\documentclass{ws-procs975x65}            
\usepackage{grffile}
\usepackage{upgreek}

\newcommand\su[1]{\mathrm{SU}(#1)}
\newcommand\Nf{N_{\textnormal{f}}}
\newcommand\mf{m_{\textnormal{f}}}
\newcommand\chit {\chi_{\textnormal{top}}}
\newcommand\chitop\chit
\def\figsubcap#1{\par\noindent\centering\footnotesize#1}

\begin{document}
\title{
\vspace*{-4cm}
{\normalsize {\rm \hfill\textnormal{LLNL-PROC-678149} }} \\
\vspace*{3cm}
Topological insights in many-flavor QCD on the lattice}

\author{Yasumichi Aoki$^a$, Tatsumi Aoyama$^a$, Ed Bennett$^{b}$, Masafumi Kurachi$^c$, Toshihide Maskawa$^a$, Kohtaroh Miura$^{ad}$, Kei-ichi Nagai$^a$, Hiroshi Ohki$^e$, Enrico Rinaldi$^f$, Akihiro Shibata$^g$, Koichi Yamawaki$^a$, Takeshi Yamazaki$^{h}$ (LatKMI Collaboration)}

\address{$^a$ Kobayashi--Maskawa Institute for the Origins of Particles and the Universe (KMI), \\Nagoya University, Nagoya, Aichi, 464-8602 Japan}
\address{$^b$ Department of Physics, Swansea University, Singleton Park, Swansea SA2 8PP UK}
\address{$^c$ Institute of Particle and Nuclear Studies, \\High Energy Accelerator Research Organization (KEK), Tsukuba, Ibaraki 305-0801 Japan}
\address{$^d$ Centre de Physique Theoretique (CPT), Aix-Marseille University, \\ Campus de Luminy, Case 907, 163 Avenue de Luminy, 13288 Marseille cedex 9, France}
\address{$^e$ RIKEN/BNL Research Center, Brookhaven National Laboratory, Upton, NY 11973-5000 USA}
\address{$^f$ Lawrence Livermore National Laboratory, 7000 East  Ave., Livermore, CA 94550-9234 USA}
\address{$^g$ Computing Research Center, High Energy Accelerator Research Organization (KEK), \\Tsukuba, Ibaraki 305-0801 Japan}
\address{$^h$ Graduate School of Pure and Applied Sciences, University of Tsukuba, \\Tsukuba, Ibaraki 305-8571, Japan}

\begin{abstract}
LatKMI Collaboration discusses the topological insights in many-flavor QCD on the lattice. We explore walking/conformal/confining phase in $\Nf=4$, $8$ and $12$ (in particular $\Nf=8$) lattice QCD via the topological charge and susceptibility, eigenvalues and anomalous dimension.
\end{abstract}


\bodymatter

\section{Introduction}

For a number of years, the LatKMI Collaboration has performed an ongoing investigation\cite{Aoki:2013zsa,Aoki:2014oha} of QCD with 4, 8, 12, and 16 light flavors on the lattice, with the aim to classify them as conformal, near-conformal, or confining and chirally broken, and calculate quantities of phenomenological relevance. In particular, $\Nf=8$ QCD is a candidate for a Walking Technicolor theory, with the potential to provide a mechanism for Electroweak Symmetry Breaking and at the same time produce a (composite) Higgs.

Thus far the study has largely been spectroscopic in nature; the aim of the work described in these proceedings was to consider orthogonal directions to gain a fuller picture of the theories. We present below first results of an investigation of the topological observables of these theories, and discuss both the effect these have on the interpretation of the spectroscopic results, and how these results may in themselves be interpreted to give signals of conformality or otherwise. Subsequently, we show some initial results of an investigation of the eigenvalue spectrum of these theories.

\section{Topological charge and susceptibility}
\begin{figure}
\begin{center}

\parbox{0.49\columnwidth}{\includegraphics[width=0.49\columnwidth]{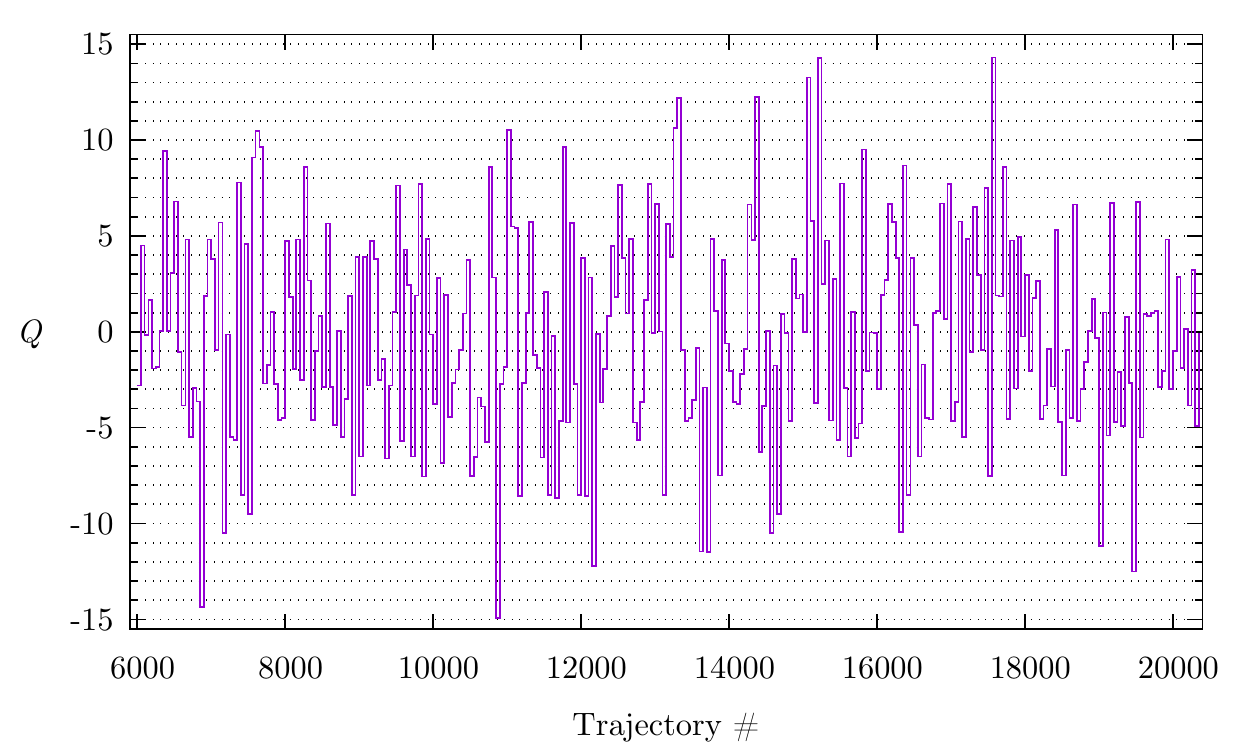}
\figsubcap{$\Nf=4,\beta=3.7,m=0.01,L=20$}}
\parbox{0.49\columnwidth}{\includegraphics[width=0.49\columnwidth]{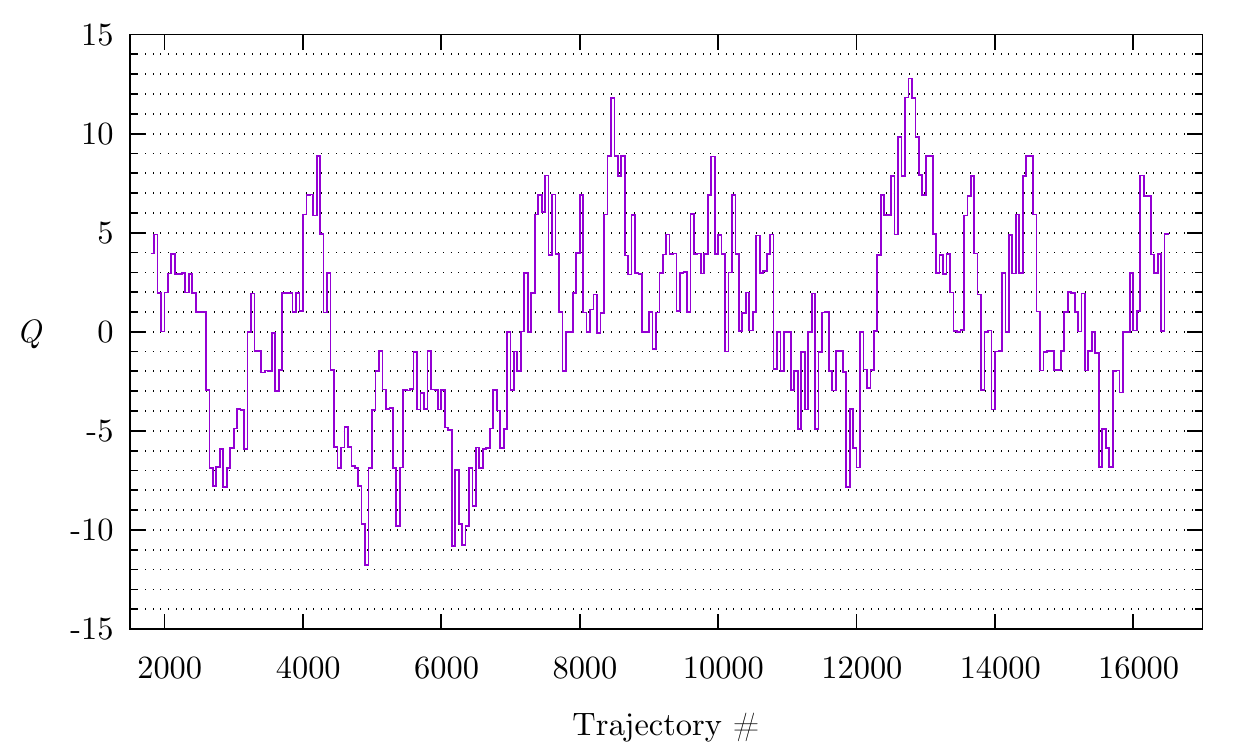}
\figsubcap{$\Nf=8,\beta=3.8,m=0.04,L=30$}}

\parbox{0.49\columnwidth}{\includegraphics[width=0.49\columnwidth]{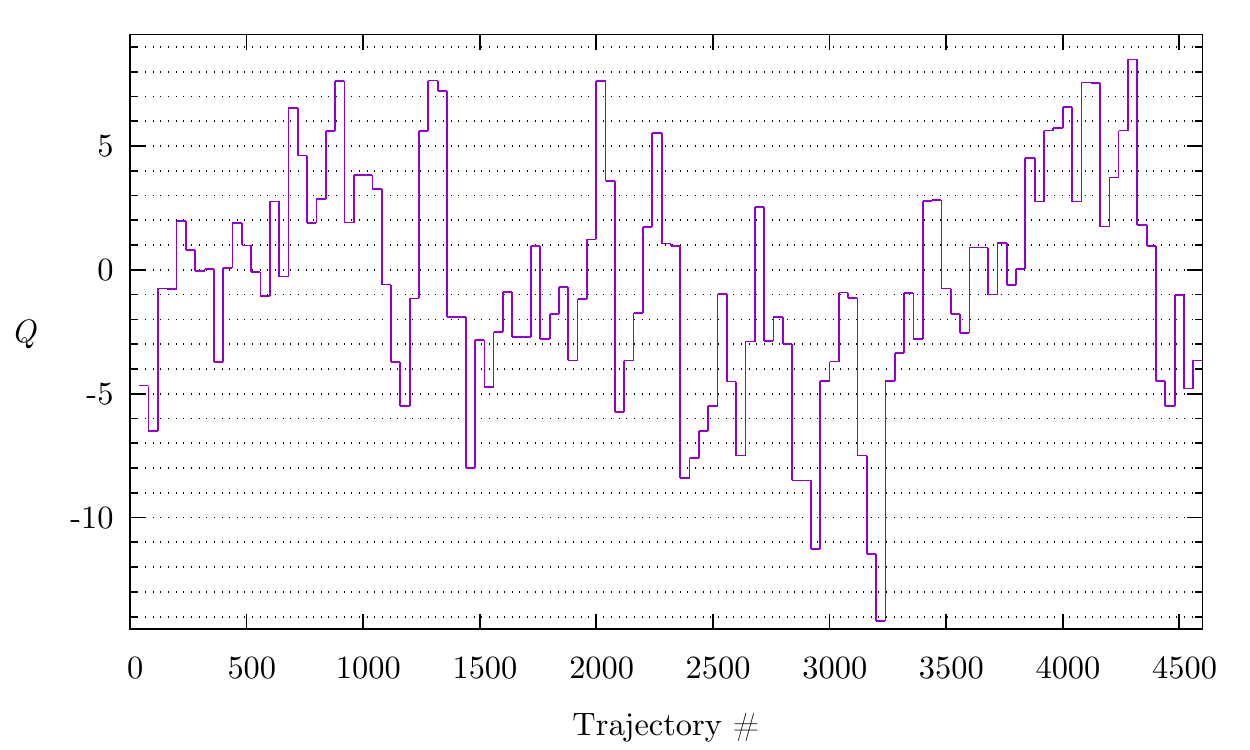}
\figsubcap{$\Nf=12,\beta=3.7,m=0.16,L=18$}}
\parbox{0.49\columnwidth}{\includegraphics[width=0.49\columnwidth]{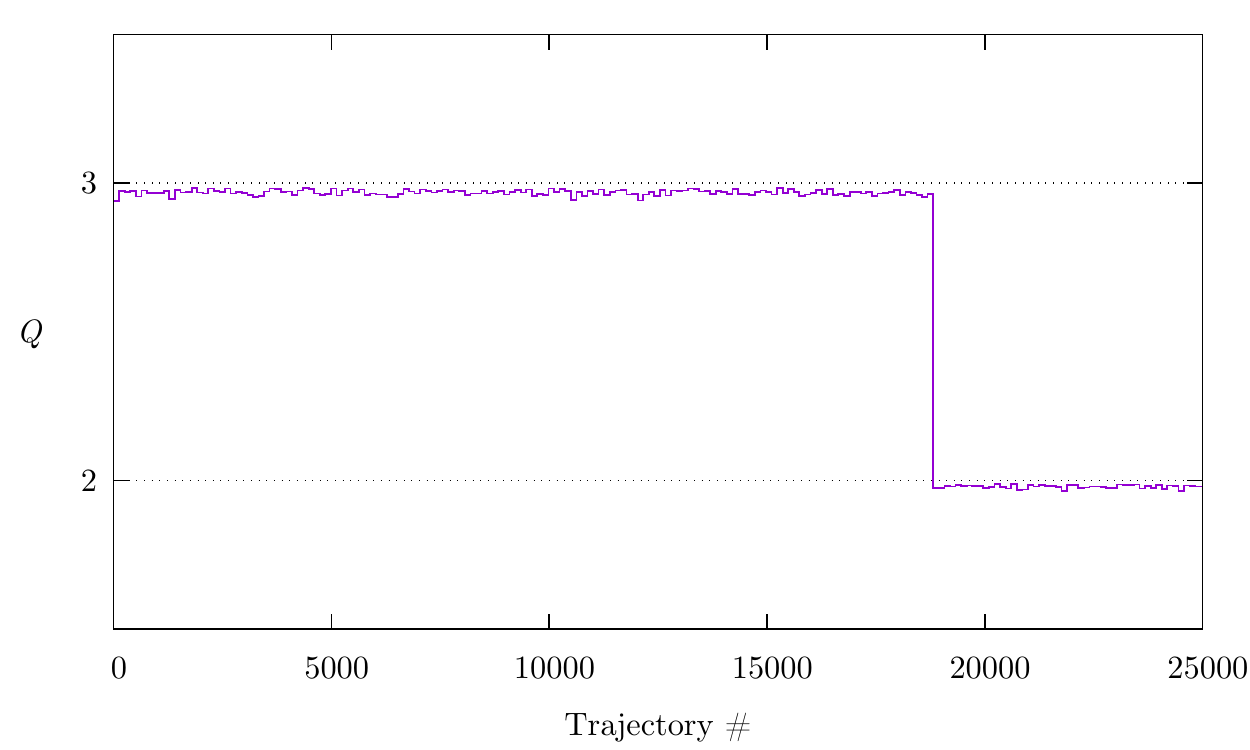}
\figsubcap{$\Nf=12,\beta=4.0,m=0.04,L=36$}}
 \caption{Sample histories of the topological charge $Q$ for the parameter sets shown. $\Nf=12,\beta=4.0$ shows the most substantial freezing observed outside of $\Nf=16$.}
\label{fig:Qhist}
\end{center}

\end{figure}

\begin{figure}
\begin{center}

\parbox{0.49\columnwidth}{\includegraphics[width=0.49\columnwidth]{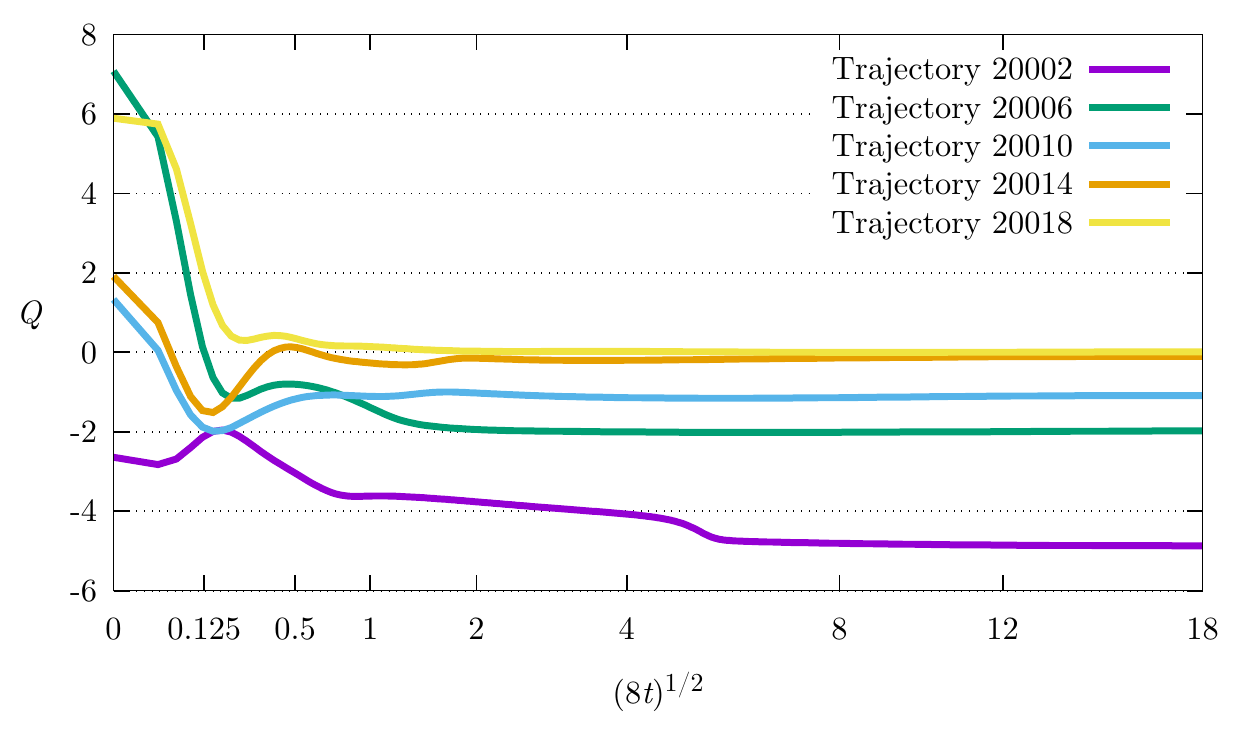}
\figsubcap{$\Nf=8,\beta=3.7,m=0.06,L=24$}}
\parbox{0.49\columnwidth}{\includegraphics[width=0.49\columnwidth]{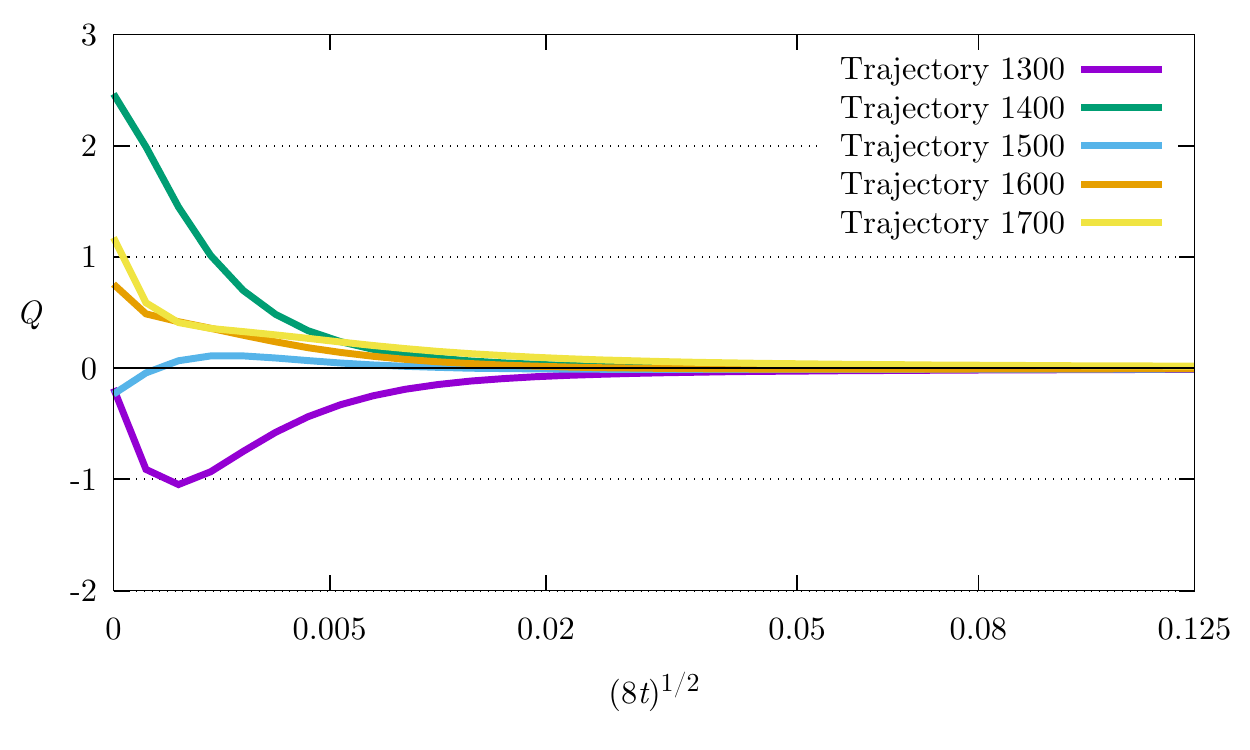}
\figsubcap{$\Nf=16,\beta=12,m=0.015,L=48$}}

\caption{History of the topological charge as a function of the gradient flow scale for selected parameter sets for $\Nf=8$ and $\Nf=16$. $\Nf=16$ is rapidly suppressed to zero, compared to $\Nf=8$ which takes longer to equilibrate and gives a variation in topological charge.}
\label{fig:flows}

\end{center}
\end{figure}

In lattice gauge theories near the continuum limit, it is possible for the topological charge $Q$ of a computation to become frozen, causing the simulation to become non-ergodic. It is thus necessary to compute the topological charge in order to verify whether or not our computations have sufficient ergodicity. Since the topological charge density operator is sensitive to UV fluctuations in the gauge field, it is necessary to smooth out the configuration. We use the gradient flow\cite{Luscher:2010iy} for this purpose; specifically the Symanzik flow. We find that $\Nf=4$ is highly ergodic, as are most parameters for $\Nf=8$, with mild loss in ergodicity emerging at the lowest values of $am$. $\Nf=12$ suffers from more ergodicity problems, with very long regions of frozen $Q$, particularly at low values of $am$. Example $Q$ histories are shown in Fig.~\ref{fig:Qhist}. Finally, the topology of $\Nf=16$ is completely suppressed, with no local minima or maxima in the topological charge density distribution indicative of the presence of a topological excitation. (An example set of cooling histories is shown in Fig.~\ref{fig:flows}. Owing to this lack of data, $\Nf=16$ will not be discussed further. 

Having calculated these values, we would like to use them for more than just a check of ergodicity. In particular, it would be useful if they could contribute to our ongoing work to study the infrared regime of these theories, specifically whether they are conformal or confining and chirally broken, and if the latter, whether they are QCD-like or exhibit walking-type dynamics. EB and Lucini have previously performed a study of this kind\cite{Bennett:2012ch} for $\su{2}$ with $\Nf=2$ adjoint flavors, and we adopt a similar procedure here. 

An IR-conformal theory becomes confining when deformed by a fermion mass. In the process, a scale is added (the fermion mass) which is intrinsically heavy---thus when considering gluonic IR observables, the theory should be indistinguishable from the equivalent quenched theory. One such observable is the topological susceptibility $\chitop$, which we calculate as 
\[
	\chitop = \frac{1}{V}\left(\langle Q^2 \rangle - \langle Q \rangle^2\right)
\]
where $V$ is the lattice volume, and $\langle Q \rangle$ is expected to be zero. Since in IR-conformal theories, the deforming mass induces non-trivial RG corrections to the lattice spacing, we must consider only dimensionless ratios. Since we ultimately want to compare the data with $\su{3}$ pure gauge theory, we scale with a gluonic observable.

\begin{figure}
\begin{center}

\parbox[b][][b]{0.42\columnwidth}{\includegraphics[width=0.42\columnwidth]{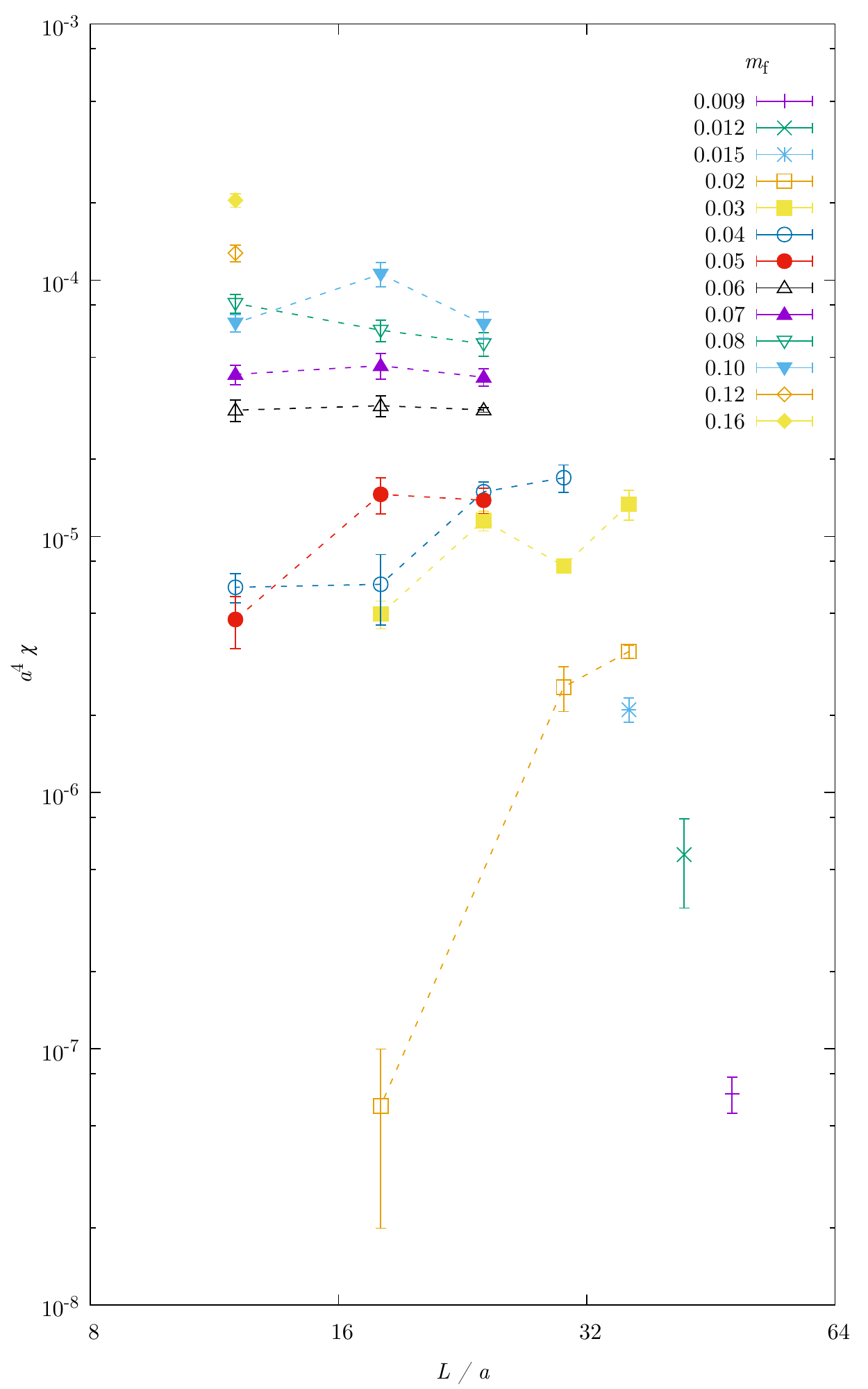}\vspace{4pt}
\figsubcap{(a)}}
\parbox[b][][b]{0.57\columnwidth}{\includegraphics[width=0.57\columnwidth]{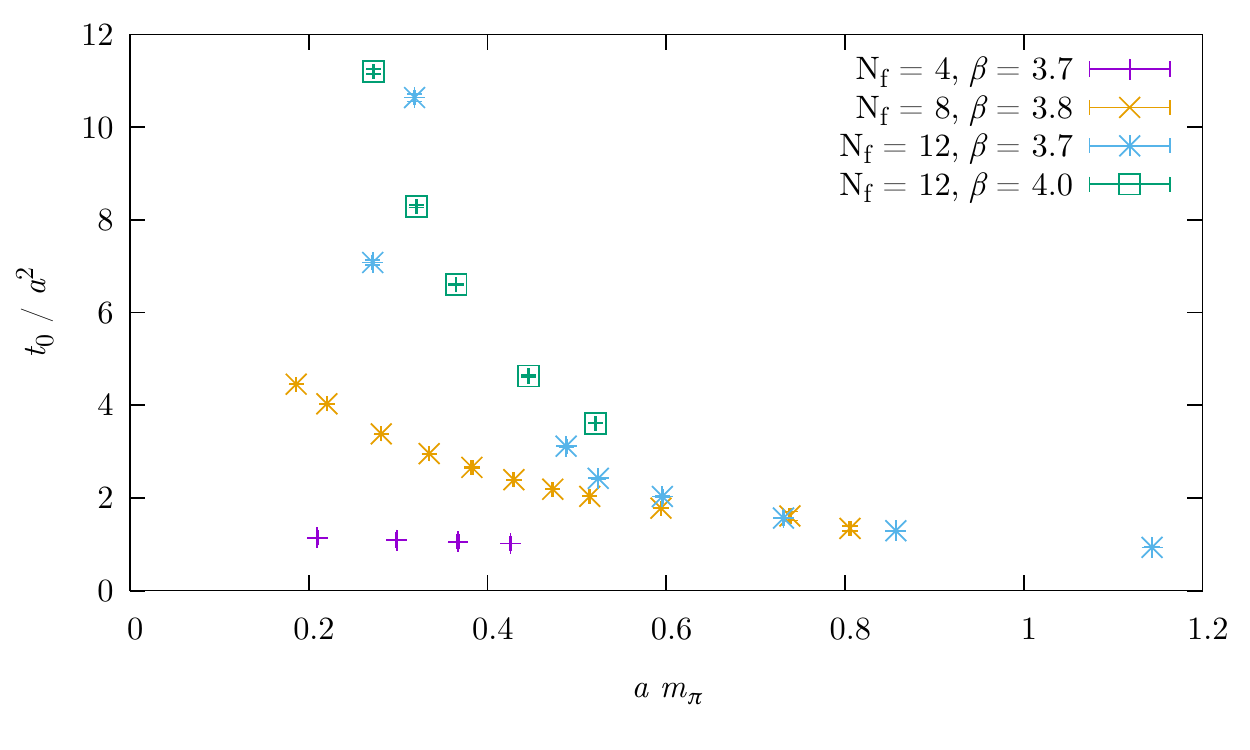}\vspace{-6pt}
\figsubcap{(b)}

\includegraphics[width=0.57\columnwidth]{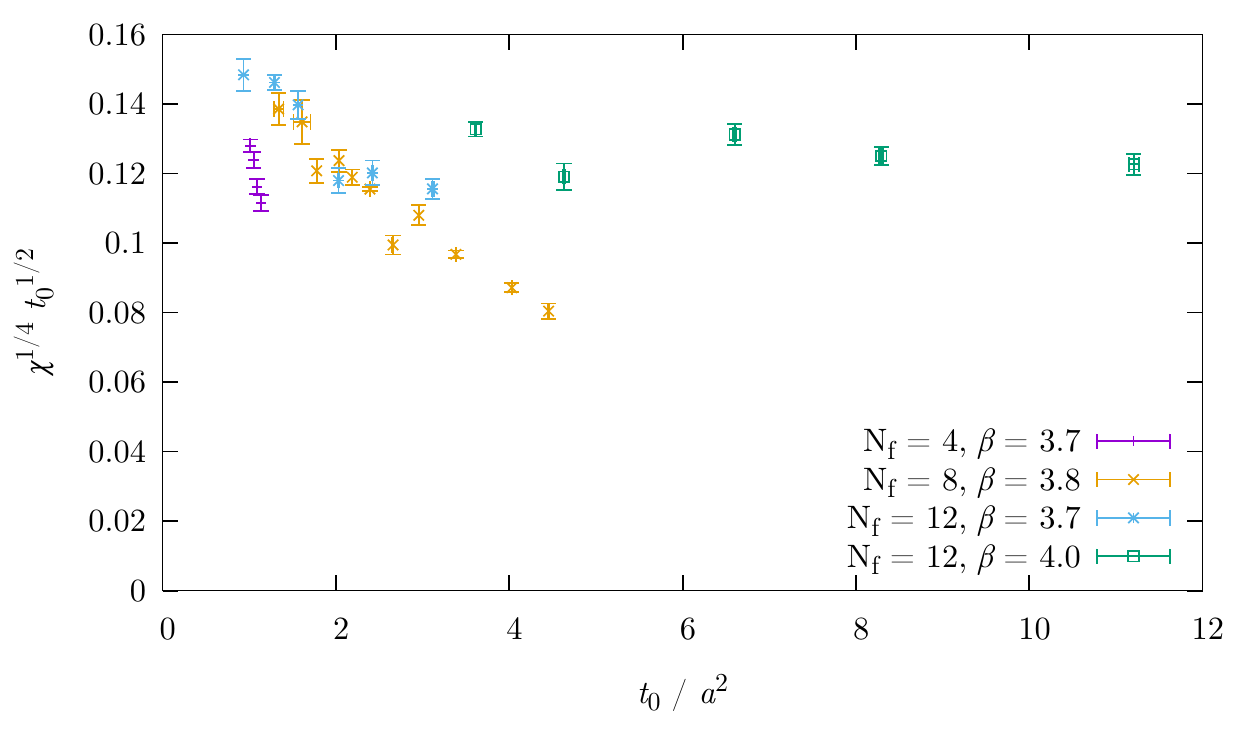}\vspace{-6pt}
\figsubcap{(c)}}

\caption{(a) Finite-volume of all available values of $\chitop$ for $\Nf=8$. (b) Comparison of the flow scale $t_0$ with the pion mass: in $\Nf=4$, the scale does not change significantly as the fermion mass changes, as we expect for a confining theory. $\Nf=12$ shows strong changes in scale as the deforming fermion mass is changed, while $\Nf=8$ lies between these two extremes. (c) Comparison of $\chitop$ scaled by the gradient flow scale $t_0$ between theories: we see that in the heavy-fermion limit (that is, as $t_0\rightarrow 0$), the theories match up. $\Nf=4$ rapidly diverges, with $\Nf=8,12$ moving away more slowly. $\Nf=12$ turns over as $t_0$ increases, which is not seen in other theories.}
\label{fig:threeplots}

\end{center}
\end{figure}

To verify that our results for the topological susceptibility are stable, we perform a finite-volume study for the $\Nf=8$ data, which is shown in Fig.~\ref{fig:threeplots}(a). In the cases suffering from topological freezing, it is obvious that the above definition of $\chitop$ will give uncontrolled errors, as we are estimating the width of the distribution based on an inadequate sampling. Here, we use the subvolume method described by the LSD Collaboration\cite{Brower:2014bqa}, to whom we defer for details of the method.


The gradient flow, having been used to access the topological charge, may also be used at no extra cost to extract a scale\cite{Luscher:2010iy} $t_0$, defined as the value of the flow time $t$ for which $t^2\langle E\rangle = 0.3$, where $E$ is a lattice discretized version of the continuum relation $E=\frac{1}{4}G_{\mu\nu}G_{\mu\nu}$. It is informative to see how this quantity varies as we move towards the chiral limit; to this end, a plot of $t_0$ against $m_\pi$ is shown in Fig.~\ref{fig:threeplots}(b). As expected for a confining QCD-like theory, $\Nf=4$ shows very little deviation in scale as $m_\pi$ varies. $\Nf=12$ varies strongly, with the scale blowing up as $m_\pi\rightarrow0$, as we expect; we observe the limit of this behaviour in $\Nf=16$, with the quantity $t^2\langle E\rangle$ flattening well before we reach $t^2 \langle E\rangle=0.3$ (not plotted here). $\Nf=8$ interpolates between these two cases; with current data we cannot distinguish whether or not $t_0/a^2$ would be finite in the limit $m_\pi\rightarrow 0$.

If we scale $\chitop$ using $t_0$, we obtain the results shown in Fig.~\ref{fig:threeplots}(c). $\Nf=4$ and $8$ show straight line behaviour as we head away from the heavy-fermion (small $t_0$ limit; $\Nf=12$ on the other hand shows a turnover. (Since the $x$-axis has a dependence on the lattice spacing, we cannot make more quantitative statements without some procedure to set the scale.)

\begin{figure}
\begin{center}

\parbox{0.49\columnwidth}{\includegraphics[width=0.49\columnwidth]{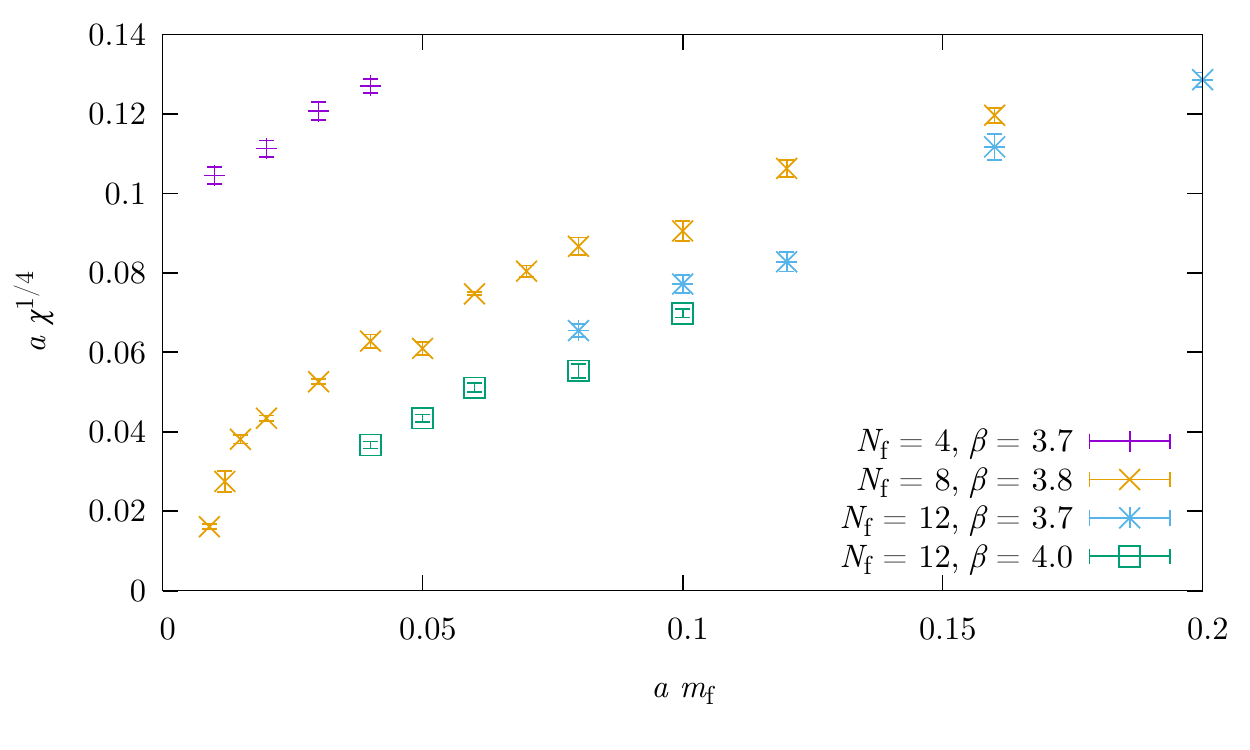}\null\vspace{-6pt}\null
\figsubcap{(a)}}
\parbox{0.49\columnwidth}{\includegraphics[width=0.49\columnwidth]{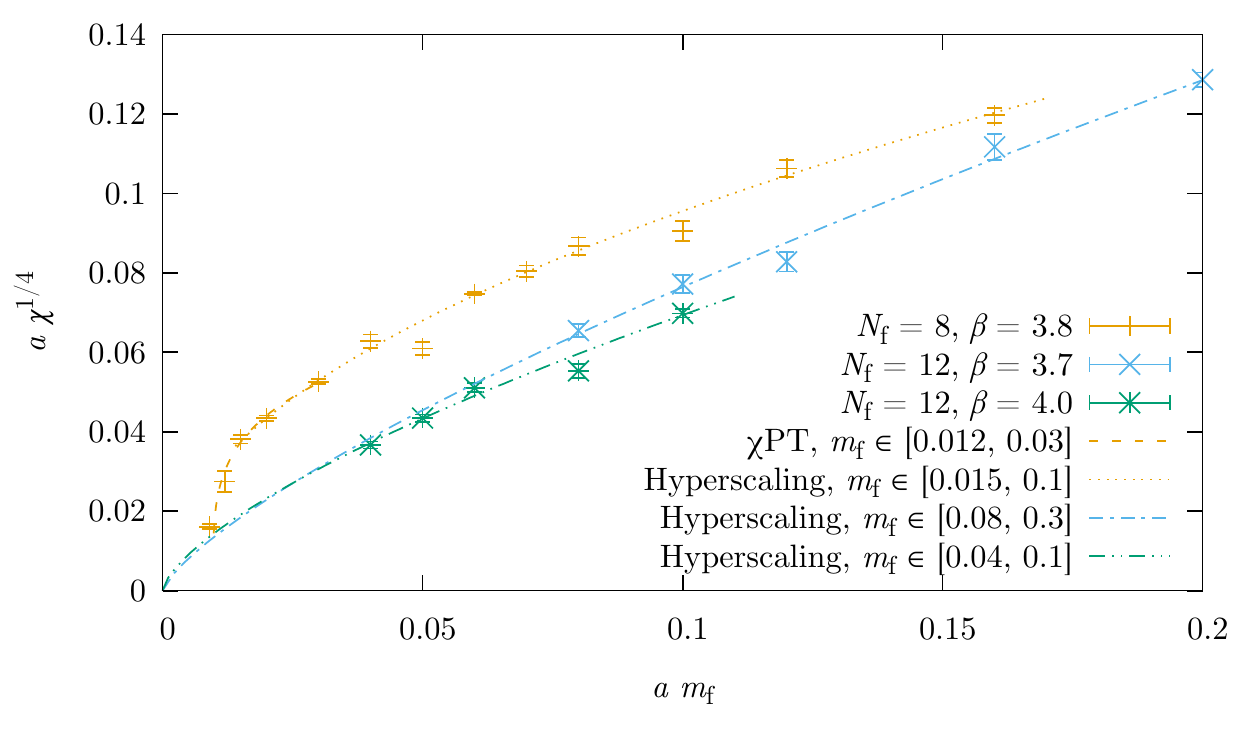}\null\vspace{-6pt}\null
\figsubcap{(b)}}

\caption{Plots of $a\chi^{1/4}$ against $\mf$, showing (a) data for $\Nf=4,8,12$, and (b) data for $\Nf=8,12$ and the results of fits performed upon these data.}
\label{fig:chi-fit}

\end{center}
\end{figure}

Finally, we look to see whether the theories obey conformal hyperscaling relations, and if so, if the chiral condensate anomalous dimension $\gamma_*$ extracted agrees with that found via other methods. In Fig.~\ref{fig:chi-fit}(a), we show the bare data for the (fourth root of) the susceptibility against the fermion mass, both in lattice units. Note that for the $\Nf=4$ data, we do not see a susceptibility of zero in the chiral limit, as we would expect for a chirally broken theory. This is most likely due to staggered taste symmetry breaking effects, which will disappear as the continuum limit is approached. This will be investigated further in a future work. 

In Fig.~\ref{fig:chi-fit}(b), we show the results of fits upon the $\Nf=8$ and $12$ data. For $\Nf=8$ we attempt both chiral perturbation theory ($\upchi$PT) 
\begin{align*}
	\chitop &= C\mf + f(a) & \Rightarrow \chitop^{1/4} &= (C\mf + f(a))^{1/4}
\end{align*}
(where $f(a)$ is a factor accounting for lattice artefacts) and hyperscaling 
\[
	\chitop^{1/4} = A\mf^{1 / (1+\gamma_*)}
\]
fits, while for $\Nf=12$ we only fit with the hyperscaling form. For $\Nf=8$, an anomalous dimension of $\gamma_*=1.04(5)$ is extracted, while for $\Nf=12$, values of 0.33(6) and 0.47(10) are found for $\beta=3.7$ and $4.0$ respectively. All values for $\gamma_*$ found are consistent with those obtained from spectroscopic studies. Both hyperscaling and $\upchi$PT fits are consistent with the data for $\Nf=8$.

\section{Eigenvalues}
\begin{figure}
\begin{center}

\hfill
\parbox{0.45\columnwidth}{\includegraphics[width=0.45\columnwidth]{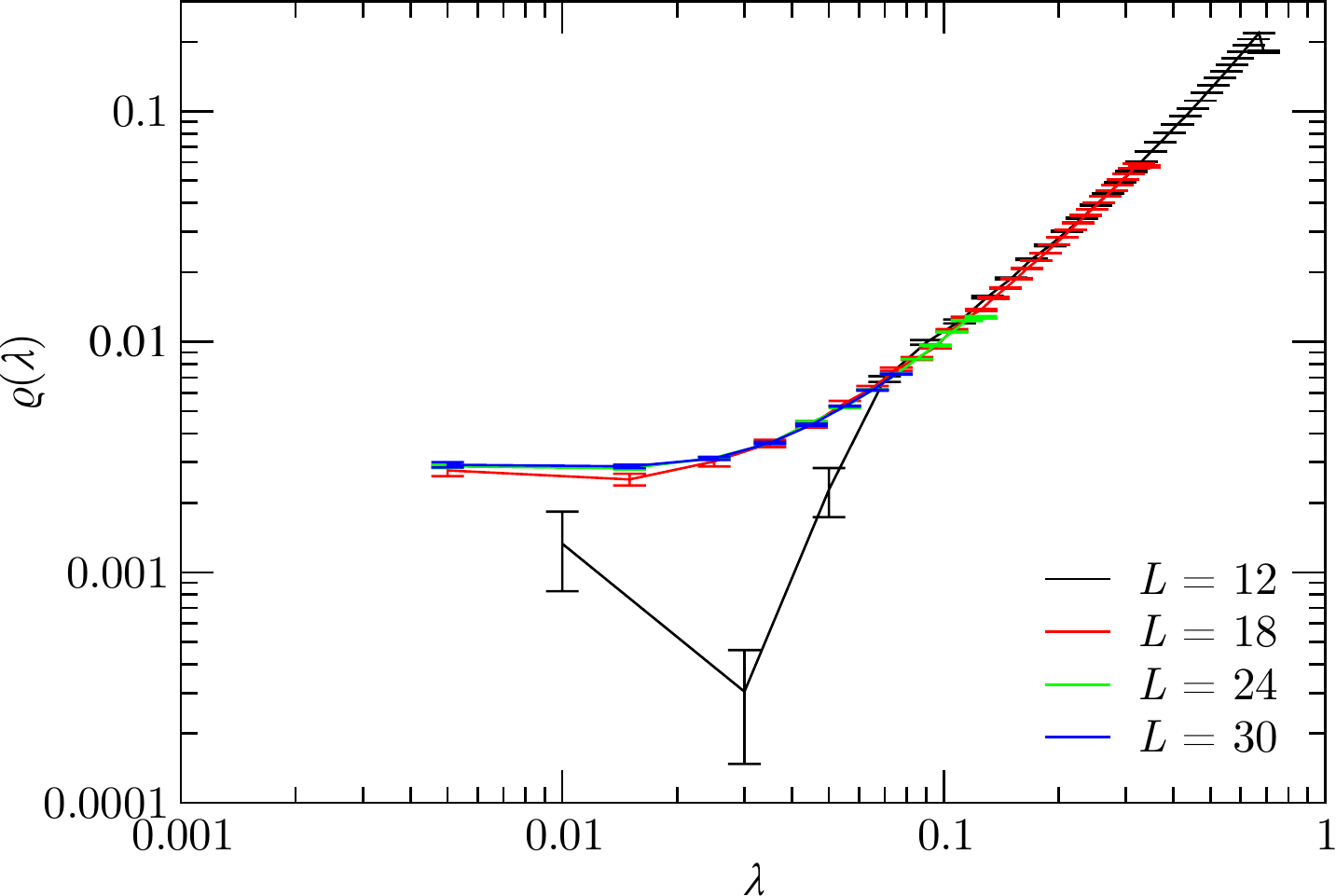}}
\hfill
\parbox{0.45\columnwidth}{\includegraphics[width=0.45\columnwidth]{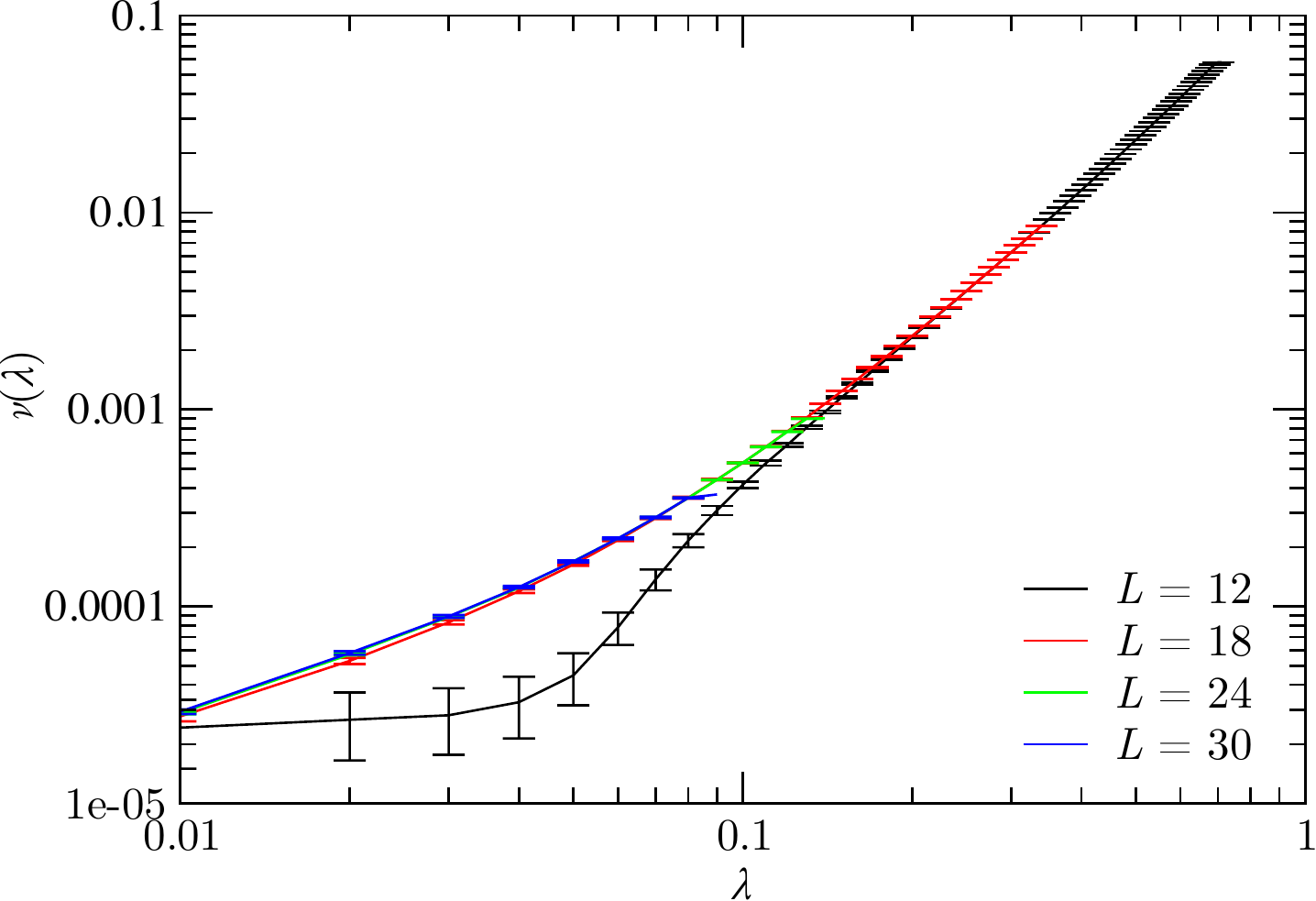}}
\hfill\null

\caption{Left: The spectral density $\rho(\lambda)$. Right: the mode number $\nu(\lambda)$, both plotted against the eigenvalue $\lambda$ as log-log plots, for $\Nf=8, \beta=3.8, m = 0.04, L=30$.}
\label{fig:rho-nu}

\end{center}
\end{figure}

The second avenue we have explored is making use of the spectrum of eigenvalues of the Dirac operator. From the Banks--Casher relation\cite{Banks:1979yr}
\[
	\langle \overline\psi \psi \rangle = \lim_{m\rightarrow 0} \lim_{V\rightarrow \infty} \pi \rho(\lambda=0)
\]
we may investigate the chiral condensate of a theory via the density of eigenvalues ($\lambda$) of the Dirac operator, $\rho(\lambda)$. However, since the eigenvalue distribution is noisy, we can integrate it to find the mode number $\nu(\lambda)$, which smooths out some noise while carrying the same information. One such distribution is plotted in Fig.~\ref{fig:rho-nu}. 

\begin{figure}
\begin{center}

\hfill
\parbox{0.45\columnwidth}{\includegraphics[width=0.45\columnwidth]{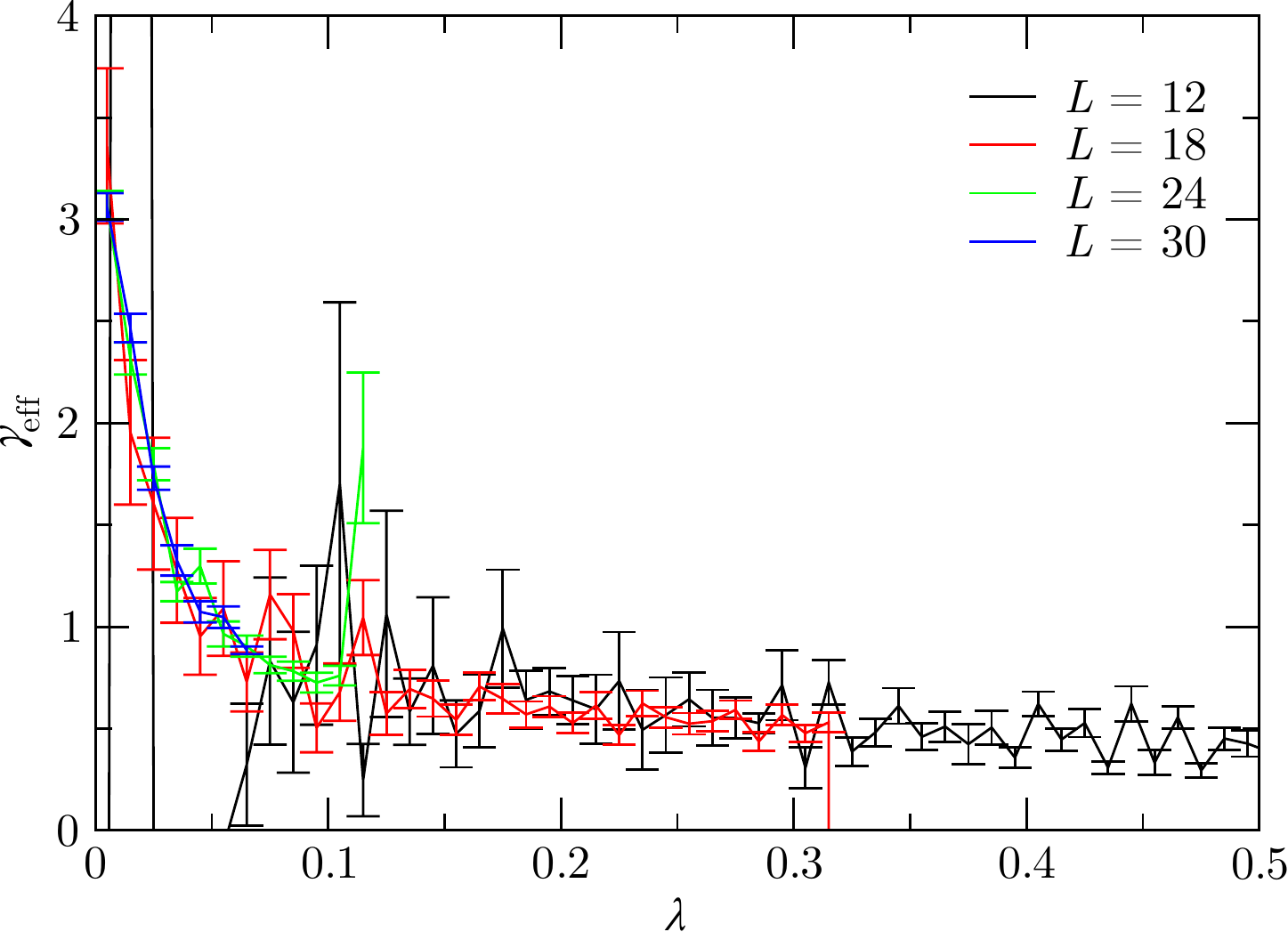}}
\hfill
\parbox{0.45\columnwidth}{\includegraphics[width=0.45\columnwidth]{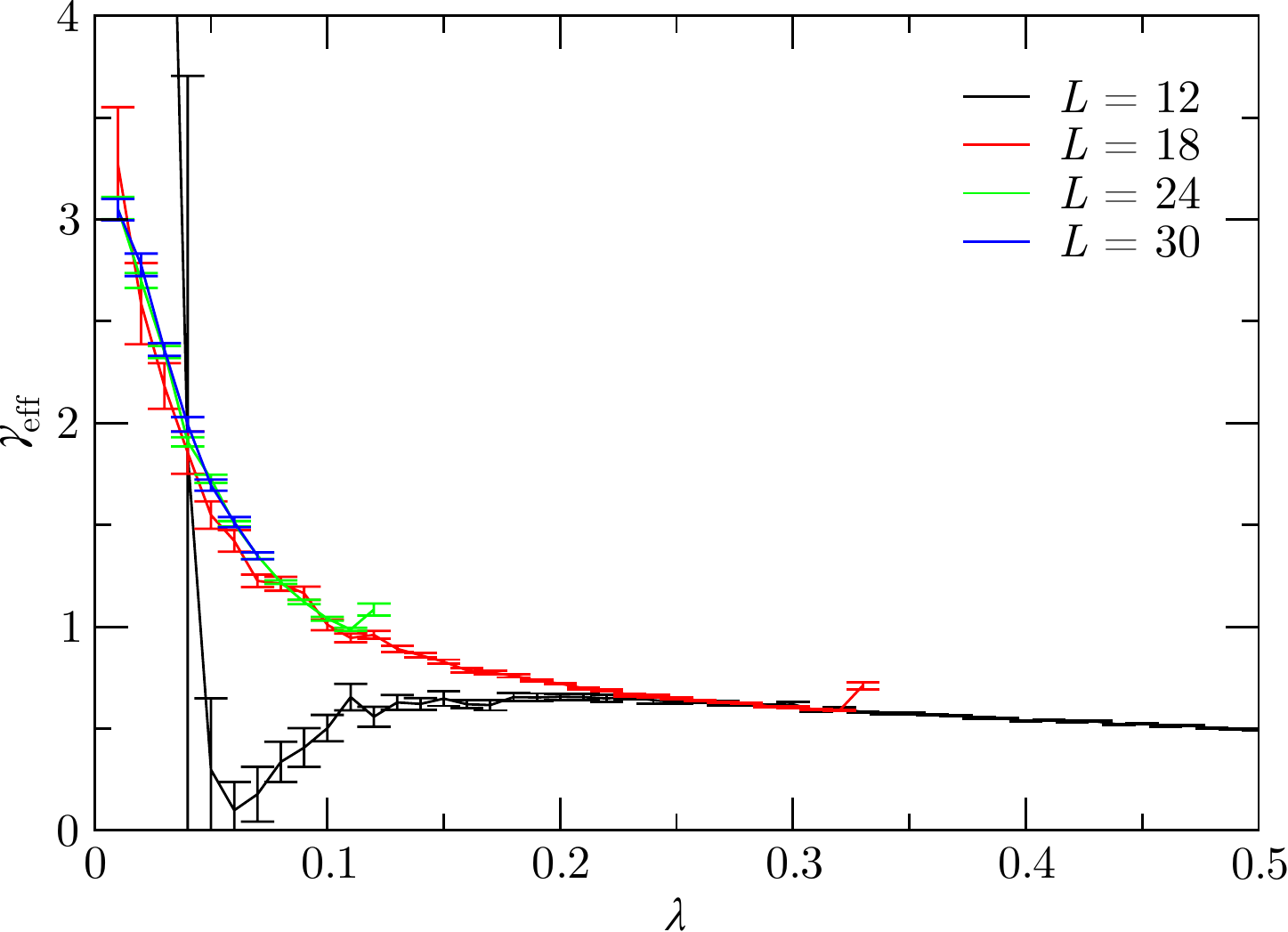}}
\hfill\null

\caption{The effective anomalous dimension $\gamma_{\textnormal{eff}}(\lambda)$, plotted against the eigenvalue $\lambda$, for the same ensemble as above. Left: via the spectral density $\rho$. Right: via the mode number $\nu$.}
\label{fig:effective-gamma}

\end{center}
\end{figure}

We would like to fit for the chiral condensate anomalous dimension. We attempt to fit using the most na\"ive form,
\begin{align*}
	\rho(\lambda) &= c\lambda^\alpha &\Rightarrow \nu(\lambda) &= d \lambda^{\alpha+1}
\end{align*}
where $\alpha+1 = 4 / (1+\gamma)$.

Since we are not in the chiral limit, we cannot simply take the limit of smallest $\lambda$; instead we define an effective exponent across an interval of $\lambda$ as
\[
	\alpha_{\textnormal{eff}}+1 = \frac{\ln \nu_2 - \ln \nu_1}{\ln \lambda_2 - \ln \lambda_1}
\]
where $\lambda_2 = \lambda_1 + \Delta$. From this we may calculate an effective anomalous dimension $\gamma_{\textnormal{eff}}$. We show the results of this fit in Fig.~\ref{fig:effective-gamma}; here we see that $\gamma_{\textnormal{eff}}(\nu)$, while having the same overall shape as $\gamma_{\textnormal{eff}}(\rho)$, has significantly less noise. All volumes appear to lie on a universal curve, save for $L=12$, which seems to suffer from finite-volume effects at small $\lambda$, while agreeing at larger values. 

By choosing a region of small $\lambda$ satisfying $\lambda\gg m$, we obtain an estimate $0.5\lesssim \gamma_{\textnormal{eff}}  \lesssim 0.7$, for the region $\lambda \gtrsim 0.2$. This value is smaller than that obtained from our spectroscopic studies, and the above fit of the topological susceptibility. The reason for this discrepancy is currently uncertain; it is possible that it is due to the finite fermion mass, or the lack of extrapolation to $\lambda=0$.

\section{Conclusions}
We have shown some initial results of our study of topological observables in $\Nf=4$, 8, and 12 QCD. We see that the topological susceptibility $\chitop$ may be used to give an estimate for the anomalous dimension of the chiral condensate $\gamma_*$. While whether $\Nf=8$ QCD is conformal or confining is still an open question, assuming a conformal ansatz gives $\gamma_*=1.04(5)$ for $\Nf=8$ at $\beta=3.8$, $\gamma_*=0.33(6)$ for $\Nf=12$ at $\beta=3.7$, and $\gamma_*=0.47(10)$ at $\beta=4.0$---consistent with estimates from spectroscopic observables. We also showed initial results for the scaling of $\chitop$ as dimensionless products with the gradient flow scale $t_0$, which at this point are inconclusive, and observed that the behavior $t_0$ itself as a function of $m_\pi$ gives some insight into the IR dynamics of the theories. The next steps of this study will be to obtain a comparable set of $\Nf=0$ data to better interpret the behaviour of $\chitop \sqrt{t_0}$, and to consider dimensionless ratios of $\chitop$ with the string tension $\sigma$.

We have also begun to study the eigenvalue spectrum of the Dirac operator, obtaining a first estimate of the anomalous dimension $0.5\lesssim \gamma_{*}  \lesssim 0.7$. Since this is inconsistent with the  results from the spectroscopy and topological susceptibility, further study is ongoing to determine why this is the case. The interaction of the gradient flow with the Dirac eigenvalues is also being investigated, as are the applications of the gradient flow to measuring the glueball spectrum and $\eta'$ mass.

\vspace{-8pt}\section{Acknowledgements}
Computations have been carried out on the $\varphi$ computer at KMI and the CX400 machine at the Information Technology Center in Nagoya University. This work is supported by the JSPS Grant-in-Aid for Scientific Research (S) No.22224003, (C) No.23540300 (K.Y.), for Young Scientists (B) No.25800139 (H.O.) and No.25800138 (T.Y.), and also by the MEXT Grants-in-Aid for Scientific Research on Innovative Areas No.23105708 (T.Y.), No. 25105011 (M.K.). E.R. acknowledges the support of the U.S. Department of Energy under Contract DE-AC52-07NA27344 (LLNL). 

\vspace{-8pt}
\bibliographystyle{ws-procs975x65}
\bibliography{ws-pro-sample}

\end{document}